**Long Range Intrinsic Ferromagnetism in Two Dimensional Materials and Dissipationless Future Technologies**

Babar Shabbir[1,2†a)], Muhammad Nadeem[3,4†], Zhigao Dai[2], Michael Fuhrer[5], Qi-Kun Xue[6], Xiaolin Wang[3], Qiaoliang Bao[2b)]

[1]Key Laboratory of Optoelectronic Devices and Systems of Ministry of Education and Guangdong Province, Shenzhen University, Shenzhen 518060, P. R. China

[2]Department of Materials Science and Engineering, and ARC Centre of Excellence in Future Low-Energy Electronics Technologies (FLEET) Monash University, Clayton, VIC 3800, Australia

[3]Institute for Superconducting and Electronic Materials, and ARC Centre of Excellence in Future Low-Energy Electronics Technologies (FLEET), University of Wollongong, Australia, NSW 2500, Australia

[4]School of Electrical Engineering and Computer Science, National University of Sciences and Technology (NUST), H-12 Islamabad, Pakistan

[5]School of Physics and Astronomy, and Monash Centre for Atomically Thin Materials and ARC Centre of Excellence in Future Low-Energy Electronics Technologies (FLEET), Monash University, Monash, VIC 3800, Australia

[6]State Key Laboratory of Low-Dimensional Quantum Physics, Department of Physics, Tsinghua University, Beijing 100084, China.

Corresponding authors: a) Babar Shabbir (Email babar.shabbir@szu.edu.cn) and b) Qiaoliang Bao (Email qiaoliang.bao@monash.edu)

[†]These authors contributed equally to this work.

## ABSTRACT

The inherent susceptibility of low-dimensional materials to thermal fluctuations has long been expected to poses a major challenge to achieving intrinsic long-range ferromagnetic order in two-dimensional materials. The recent explosion of interest in atomically thin materials and their assembly into van der Waals heterostructures has renewed interest in two-dimensional ferromagnetism, which is interesting from a fundamental scientific point of view and also offers a missing ingredient necessary for the realization of spintronic functionality in van der Waals heterostructures. Recently several atomically thin materials have been shown to be robust ferromagnets. Such ferromagnetism is thought to be enabled by magneto crystalline anisotropy which suppresses thermal fluctuations. In this article, we review recent progress in two-dimensional ferromagnetism in detail and predict new possible two-dimensional ferromagnetic materials. We also discuss the prospects for applications of atomically thin ferromagnets in novel dissipationless electronics, spintronics, and other conventional magnetic technologies. Particularly atomically thin ferromagnets are promising to realize time reversal symmetry breaking in two-dimensional topological systems, providing a platform for electronic devices based on the quantum anomalous Hall Effect showing dissipationless transport. Our proposed directions will assist the scientific community to explore novel two-dimensional ferromagnetic families which can spawn new technologies and further improve the fundamental understanding of this fascinating area.

**TABLE OF CONTENTS**

# I. INTRODUCTION

Ferromagnetism (FM) is a physical phenomenon realized in certain materials that retain a permanent magnetic moment in the absence of any external magnetic field as long as the temperature is below a critical value, generally known as the Curie temperature "$T_c$". Ferromagnetic materials have long been recognized for their technological importance, beginning with the compass, and are still essential today in electric motors and generators and non-volatile information storage. While ferromagnetism in three dimensions (3D) can be achieved with $T_c$ much greater than room temperature, the spontaneous rotational symmetry breaking associated with isotropic ferromagnetism is strictly forbidden in two dimensions (2D). While certain anisotropic models are known to exhibit stable ferromagnetism in 2D, it is generally recognized that the increased instability of 2D materials to thermal fluctuations poses challenges to realizing ferromagnetism with high $T_c$.

The past one and a half decades have seen an explosion of interest in atomically thin materials since the isolation of graphene, an atom-thick plane of carbon[1]. This class of materials is also known as van der Waals materials due to the weak bonding between layers, making these materials compatible with a variety of substrates and also opening a vast new field of "van der Waals heterostructures" which combine different atomically thin materials to form new structures with new properties[2]. Atomically thin materials have been discovered with a wide range of properties: metals, semiconductors, insulators, superconductors, charge-density-wave materials, and ferroelectrics. Ferromagnetism was notably absent from this list until recent discoveries of atomically thin ferromagnets $Cr_2Ge_2Te_6$[3], $CrI_3$[4], and $VSe_2$[5]. The discovery of intrinsic ferromagnetism in 2D materials is vital for understanding spin behavior in low dimensions and could enable novel spintronics applications ranging from molecular quantum devices and sensing to high-density data storage ultrathin devices[6-9]. 2D ferromagnets may also be essential components in proposed heterostructures which can exhibit the quantum anomalous Hall effect at high temperatures, highly desirable for applications in dissipationless electronics and spintronics.

Since the discovery of graphene[10], several techniques have been proposed to extrinsically induce long-range magnetic order into 2D materials such as defect engineering[11], absorption of magnetic ions[12] or proximity effect[13]. In defect engineering, the simplest strategy for realizing magnetism in non-magnetic materials is to create unpaired electrons by modifying its electronic structure via vacancies[14, 15], adatoms[12], grain boundaries[16] or edges[17]. In absorption technique, magnetic ions are absorbed inside 2D vdW materials where magnetic order emerges when the magnetic ions experience exchange coupling. On the other hand, in magnetic proximity effect, 2D vdW materials experience exchange coupling when placed in contact with ferromagnetic insulating substrates.

Defect engineering and absorption induced ferromagnetic systems usually come along with a change in their crystalline structure. The obtained small magnetic moments in such kind of disordered systems, the need to control the doping/adsorption precisely, and/or the sacrifice of high carrier mobilities due to scattering are not suitable for hosting robust exchange interaction between extrinsically introduced local magnetic moments. Moreover, apart from the debated experimental evidence for defect engineering and absorption techniques, proposed theoretical models for inducing low density ferromagnetism by tuning the band structure[18-20] have yet to be achieved experimentally.

On the other hand, proximity induced ferromagnetic systems have more advantages and promise preservation of the original crystalline structure, better controllability of the electronic states, uniform magnetization, and uniformly distributed band gap in space, etc. However, proximity induced magnetic responses are limited in 2D materials. These potential scarce in externally induced ferromagnetism motivates to find ferromagnetism originating intrinsically from the parent 2D lattice.

However, little attention has been drawn to investigate the intrinsic ferromagnetism in two-dimensional "2D" materials because the long-range magnetic order in 2D systems could be destroyed by thermal fluctuations, for example the Mermin–Wagner theorem indicates that long-range magnetic order is impossible at any finite temperature in an isotropic 2D spin system[21-23]. To suppress the thermal fluctuations and achieve long range intrinsic ferromagnetism in 2D materials, magnetic anisotropy is necessary. Based on various forms of magnetic anisotropies, recent groundbreaking experimental realizations of intrinsic ferromagnetism in $Cr_2Ge_2Te_6$, $CrI_3$, and especially $VSe_2$ at room temperature have been achieved. Surprisingly, monolayer $VSe_2$ has been revealed to be ferromagnetic while 3D $VSe_2$ is paramagnetic, pointing out that mono-layers or two-dimensional (2D) structures can reveal distinctive properties in contrast to their multi-layers or three dimensional (3D) structures.

This review focuses on intrinsic ferromagnetism in 2D materials. By "intrinsic ferromagnetism", we intend to limit the discussion to ferromagnetism occurring in a stoichiometric compound or heterostructure of stoichiometric compounds, as opposed to defect or dopant induced ("dilute") ferromagnetism. By "2D material" we intend those materials which can be stably isolated in few atom thick layers; typically these materials consist of covalently bonded layers which interact with each other through weaker van der Waals bonding.

This paper also explores the role of 2D ferromagnetism in various topological phases of quantum matter. The quantum anomalous Hall effect, which enables robust dissipationless electronic transport over macroscopic distances, is entangled with long-range intrinsic ferromagnetism in 2D materials[24, 25]. The realization of quantum anomalous Hall effect through magnetically doped topological insulators will be discussed within the Landau framework of phase transitions, especially emphasizing time reversal symmetry breaking and the associated ferromagnetic order.

Finally, this timely review will help the community to discover novel magnetic materials that can lead to ultrathin 2D devices which will be more efficient, compact, and could be governed by using precise and reliable systems. Due to many available reviews about ferromagnetism in 3D and/or their other associated phenomena in literature[9, 23, 26-28], this review focuses solely on intrinsic ferromagnetism in 2D materials, and especially draws attention to proposed materials, i.e. which are mainly predicted by density functional theory (DFT) calculations, as well as experimentally realized 2D materials and future directions.

## II. ROLE OF MAGNETIC ANISOTROPY IN 2D FERROMAGNETIC MATERIALS

Magnetic anisotropy is the dependence of the magnetic properties on the relative orientation of the applied magnetic field with respect to the crystal lattice. Magnetic anisotropy is an essential property of magnetic materials, having significance ranging from fundamental to technological, for example, magnetic anisotropy is essential for aligning the magnetic moments in magnetic storage media. Magnetic anisotropy is also a prerequisite to observe ferromagnetism in 2D materials[29]. There are different types of magnetic anisotropy such as shape anisotropy, exchange anisotropy, magnetoelastic anisotropy and most importantly magnetocrystalline anisotropy[29]. Magnetocrystalline anisotropy is caused by the spin-orbit coupling which couples the spin orientation to the orientation of the electron orbitals[30]. The structure becomes important when the average magnetization of a specimen as a function of an applied external field is considered since the magnetization depends on the orientation of the magnetic field with respect to the crystal axes. Additionally, external factors such as pressure which is commonly used to tune magnetic properties[31-35] can also modify the magnetic anisotropy[36].

While magnetically isotropic materials can exhibit long range ferromagnetic order in 3D, such ordering is prohibited in 2D due to the thermal fluctuations. To achieve long-range ferromagnetism in 2D systems, magnetic anisotropies need to be incorporated with exchange interaction:

$$H = -\frac{1}{2}\sum_{i,j} J_{ij}\mathbf{S}_i \cdot \mathbf{S}_j - \sum_i A_i (S_i^z)^2 - g\mu_B B \sum_i S_i^z \qquad (1)$$

where the first term represents Heisenberg isotropic model, second term represents magnetocrystalline anisotropy whereas the last term is the interaction of local moments with an external magnetic field $B$. Here $S_i$ is the spin operator on site $i$, $J_{ij}$ is the exchange coupling between spins on site $i$ and $j$ whose positive and negative values represent ferromagnetic and antiferromagnetic ordering, $A$ is the single-ion anisotropy, whereas $g$ and $\mu_B$ are landau g-factor and Bohr magneton respectively.

Cheng Gong et al.,[3] showed recently that long-range ferromagnetic order persists in a few layers of a $Cr_2Ge_2Te_6$ if magnetocrystalline anisotropy and Zeeman effect are considered along with Heisenberg exchange interaction. Similarly, Bevin Huang et al.,[4] also showed that ferromagnetic order persists even in a monolayer of chromium tri-iodide $CrI_3$ if magnetocrystalline anisotropy and Zeeman effect are incorporated with Ising exchange interaction. Recently J. L. Lado et al. [37] showed that magnetic anisotropy in $CrI_3$ mainly originates from the superexchange interaction across the Cr-I-Cr bonds, due to the especially large spin orbit coupling of iodine, with a small additional contribution due to the single-ion anisotropy of Cr. The large magnetic anisotropy due to superexchange can stabilize the magnetic ordering in 2D $CrI_3$.

Similarly, Bonilla et al., showed that M–H loop VSM measurements of the $VSe_2$ monolayer /HOPG or $VSe_2$ monolayer /$MoS_2$ sample with 1T metallic phase indicate strong in-plane anisotropy. This result is consistent with the predictions made through first principle simulations that both semiconducting 2H and metallic 1T phases of monolayer transition-metal dichalcogenides exhibit ferromagnetic ordering with in-plane magnetization direction[38-40]. This strong magnetic anisotropy resulting in-plane spin polarization in transition-metal dichalcogenide monolayer mainly originates from both through-bond and through-space interactions between transition metal anions and chalcogenide cations[39].

# III. PROPOSED 2D FERROMAGNETIC MATERIALS AND THEIR EXPERIMENTAL REALIZATION

Figure 1(A) shows a graph for various proposed 2D ferromagnetic materials with their respective $T_c$, which were mainly predicted via DFT calculations. The 2D ferromagnetism is mostly predicted in transition metal chalcogenides and transition metal halides, therefore, their general crystal structures including top and side views are presented in Fig 1(B). Transition metal trichalcogenides (TMT) are compounds of the formula $ABX_3$ where A is a 3d metal (Cr), B can be metalloid or metal (such as Si, Ge, Sn) and X is a chalcogen (S, Se, Te)[41-45]. Generally, the transition metal atoms form a honeycomb lattice structure within each layer, and each of them is octahedrally coordinated by six chalcogenide atoms from its three neighboring ($B_2X_6$) ligands, with the centers of the hexagons occupied by the $B_2$ groups[41]. Magnetic order is predicted in the single-layer limit of these van der Waals materials, and the configuration of the magnetic phase is sensitive to the transition metal/chalcogen element combinations. Likewise, transition metal dichalcogenides (TMD) are 2D materials of the formula $AX_2$, with A a transition metal (e.g., V, Cr, and Mn) and X a chalcogen (e.g. S and Se)[46-48]. Ferromagnetism is predicted in T phase TMDs, where each transition metal adopts an octahedrally coordinated structure with respect to the six chalcogenide atoms around it and each chalcogenide atom is bonded to three nearby transition metal atoms[46]. The localized magnetic moments on the transition metal atoms are coupled ferromagnetically by the super-exchange interactions mediated by the chalcogenide atoms. Trihalides (TH) are compounds of formula $AZ_3$ where A is a transition metal (*e.g.* Ru, Fe) and Z a halogen (*e.g.* I, Br)[49-52]. The crystal structure of TH consists of three flat atomic layers i.e. halide-transition metal-halide from top to bottom layers, with transition metal atom arranged in a honeycomb lattice, each coordinated to six halide atoms[52]. The strong spin exchange interaction and huge magnetic anisotropy play a vital role to observe FM in 2D systems. Dihalides (DH) are compounds of formula $AZ_2$ with A a transition metal and Z a halogen[53, 54]. Generally, a layer of A atoms arranged in a triangular lattice is sandwiched between two layers of Z atoms in triangular lattices, similar to the TMDs, with the 1T crystal structure ($C_{3v}$ symmetry) preferred in metal dihalide families. Each metal atom is surrounded by six neighboring halogen atoms forming an octahedral $[AX_6]^{-4}$ unit. The magnetic states in DH evolve from the competition between anti-ferromagnetic direct nearest-neighbor *d–d* exchange and ferromagnetic super-exchange *via* halogen *p* states[53]. MXenes are 2D inorganic compounds, composed of few atom thick layers of transition metal carbides, nitrides, or carbonitrides. Several combinations of MXenes; i.e., $M_2AX_2$ where M = (Mn, Cr, Ti), A= (C, N), X= (O, OH, F, Cl); and $A_{n+1}X_n$ with A= Cr, X= (C, N); are ferromagnetic in nature[55-59]. Here, the superexchange mechanism contributes to the magnetism. Furthermore, ferromagnetism has also been predicted in several other materials (denoted by OM in Table 1) including materials such as $MnO_2$, $FeC_2$ etc[60-71]. It is also important to mention here that Mounet and co-workers have also theoretically identified a group of ferromagnetic metals and semiconductors using data-mining algorithms and advanced, high-throughput electronic structure theory[72]. Table 1 shows a list of predicted materials with their curie temperatures ($T_c$). In theoretical studies, $T_c$ values are mostly determined by using a mean field theory (MFT) approach or statistical Monte Carlo simulations (MC). The $T_c$ values predicted by MFT are generally over estimated[73, 74]. On the other hand, more accurate $T_c$ values can be computed using statistical Monte Carlo (MC) simulations based on the Ising model[43].

(A)

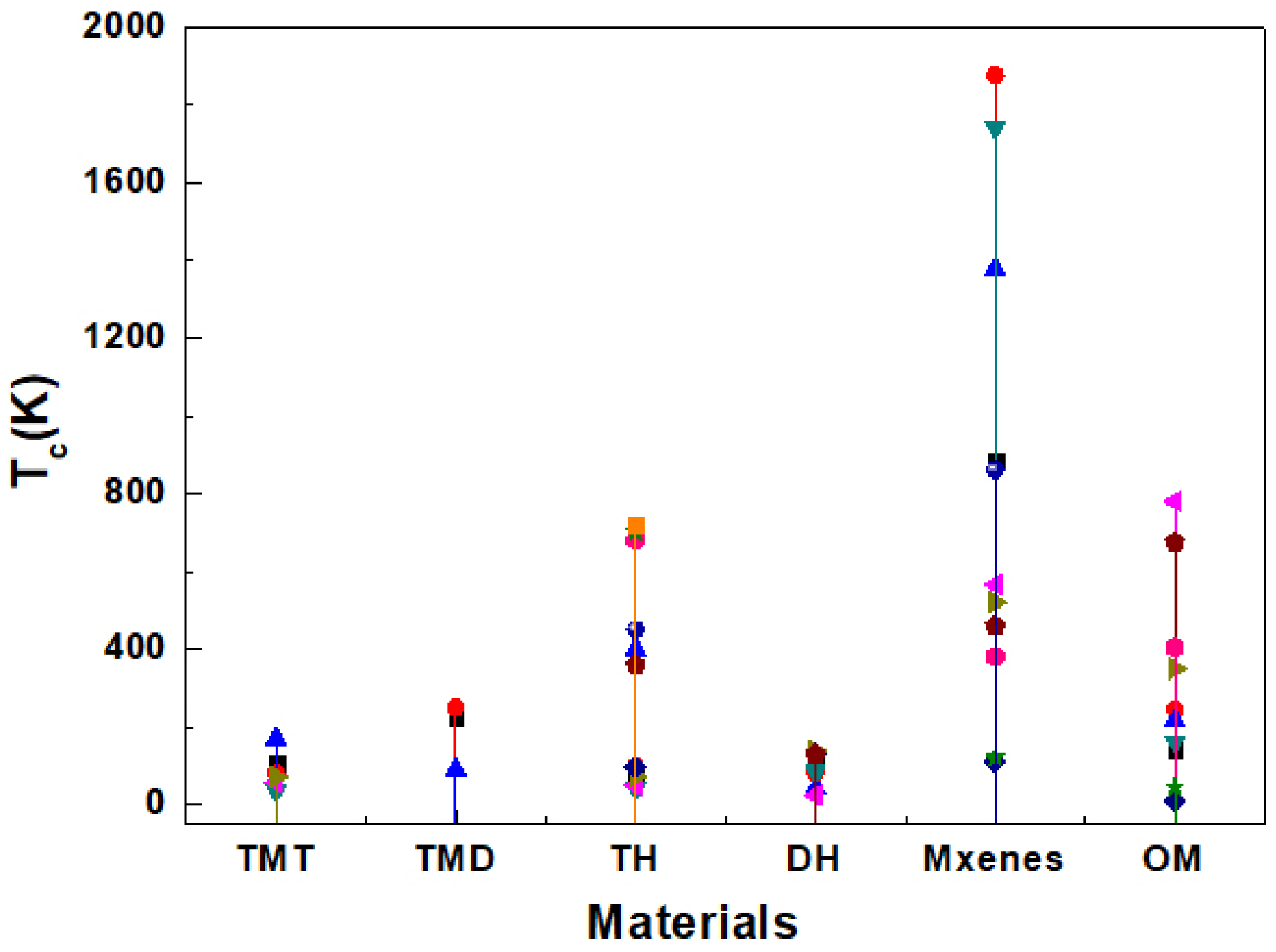

(B)

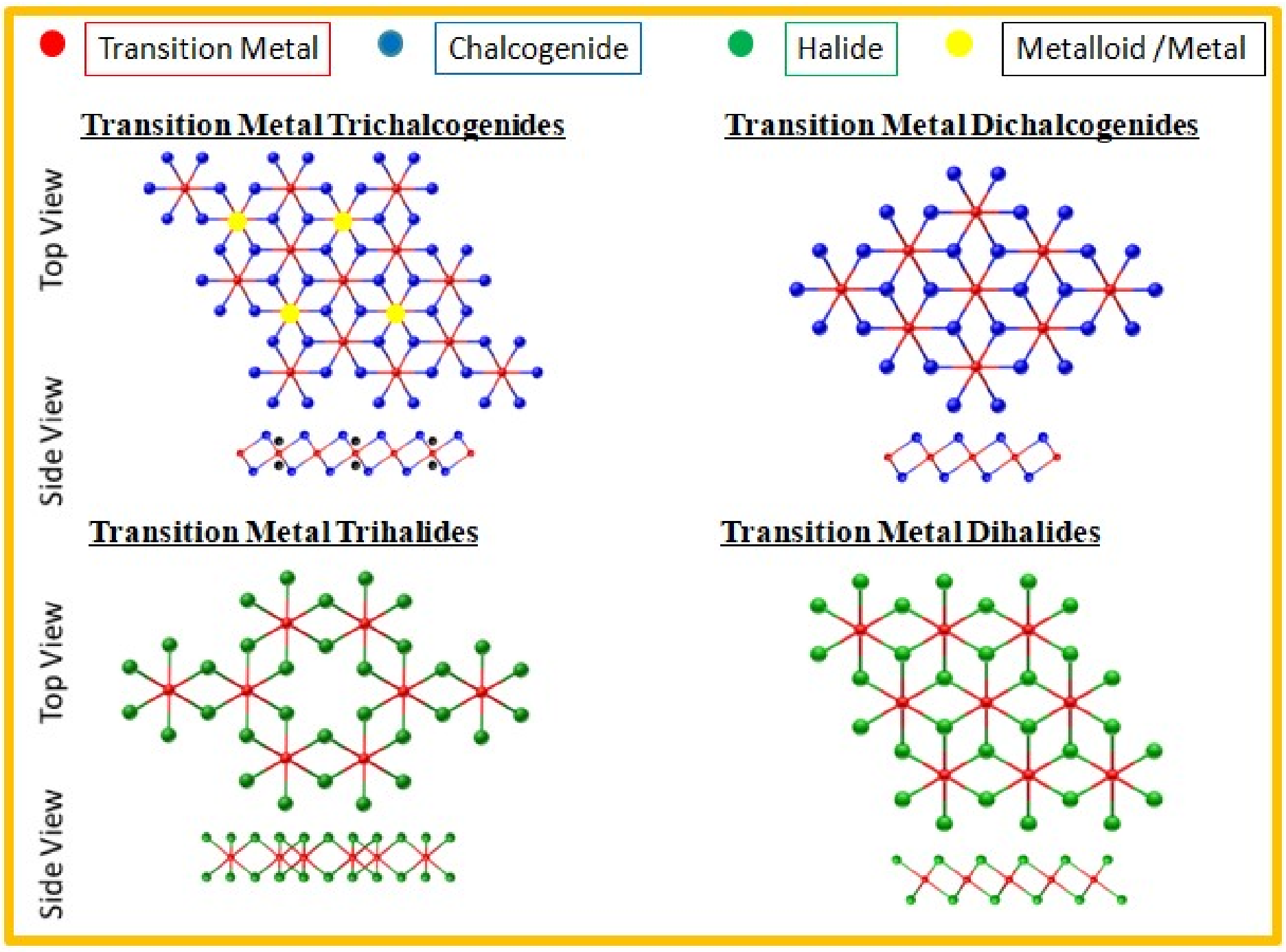

**Figure 1(A):** Proposed 2D ferromagnetic materials with their respective Curie Temperatures "$T_c$", determined via DFT calculations ($T_c$ details are given in Table 1). TMT (Transition Metal Trichalcogenides)[41-45], TMD (Transition Metal Dichalcogenides)[46-48, 75], TH (Tri-Halides)[49-52, 76], DH (Di-Halides)[53], MXenes[55-59], OM (Other materials) [60-71]. **(B):** Crystal structures of TMT (Transition Metal Trichalcogenides), T Phase-TMD (Transition Metal Dichalcogenides), TH (Transition Metal Trihalides) and DH (Transition Metal Dihalides).

**Table 1:** Materials and their corresponding Curie temperatures $T_c$(K). Materials listed in black box are experimentally realized ferromagnets. TMT (Transition Metal Trichalcogenides), TMD (Transition Metal Dichalcogenides), TH (Tri-Halides), DH (Di-Halides), MX (MXenes), OM (Other materials), SCs (Semiconductors) and HMs (Half Metals)

| Material | Tc(K) | Class | Nature | Reference | Material | Tc(K) | Class | Nature | Reference |
|---|---|---|---|---|---|---|---|---|---|
| $CrGeTe_3$ | 106 | TMT | SCs | 41 | $VCl_3$ | 80 | TH | HMs | 49 |
| $CrSiTe_3$ | 80 | TMT | SCs | 42 | $VI_3$ | 98 | TH | HMs | 49 |
| $CrSnTe_3$ | 170 | TMT | SCs | 43 | $NiCl_3$ | 400 | TH | Insulator | 50 |
| $CrSiTe_3$ | 36 | TMT | SCs | 44 | $CrF_3$ | 41 | TH | SCs | 51 |
| $CrGeTe_3$ | 57 | TMT | SCs | 44 | $CrCl_3$ | 49 | TH | SCs | 51 |
| $CrGaTe_3$ | 71 | TMT | SCs | 45 | $CrBr_3$ | 73 | TH | SCs | 51 |
| $MnS_2$ | 225 | TMD | SCs | 46 | $CrI_3$ | 95 | TH | SCs | 51 |
| $MnSe_2$ | 250 | TMD | SCs | 46 | $RuI_3$ | 360 | TH | Insulator | 52 |
| $CrS_2$ | -- | TMD | SCs | 48 | $MnCl_3$ | ~680 | TH | HMs | 76 |
| $VS_2$ | 90 | TMD | SCs | 47 | $MnBr_3$ | ~700 | TH | HMs | 76 |
| $Cr_3C_2$ | 886 | MX | HMs | 56 | $MnF_3$ | 450 | TH | HMs | 76 |
| $Mn_2NF_2$ | 1877 | MX | HMs | 58 | $MnI_3$ | 720 | TH | HMs | 76 |
| $Mn_2NO_2$ | 1379 | MX | HMs | 58 | $FeCl_2$ | 109 | DH | HMs | 53 |
| $Mn_2N(OH)_2$ | 1743 | MX | HMs | 58 | $FeBr_2$ | 81 | DH | HMs | 53 |
| $Cr_2NO_2$ | 566 | MX | HMs | 58 | $FeI_2$ | 42 | DH | HMs | 53 |
| $Mn_2CF_2$ | 520 | MX | HMs | 59 | $CoCl_2$ | 85 | DH | -- | 53 |
| $Cr_2C$ | -- | MX | HMs | 55 | $CoBr_2$ | 23 | DH | -- | 53 |
| $Cr_2CF_2$ | -- | MX | Metal | 57 | $NiCl_2$ | 138 | DH | SCs | 53 |
| $Cr_2C(OH)_2$ | -- | MX | Metal | 57 | $NiBr_2$ | 132 | DH | -- | 53 |
| $Cr_2NF_2$ | -- | MX | Metal | 57 | $NiI_2$ | 129 | DH | -- | 53 |
| $Cr_2N(OH)_2$ | -- | MX | Metal | 57 | $MnO_2$ | 140 | OM | SCs | 62 |
| $Cr_2NO_2$ | -- | MX | Metal | 57 | $FeC_2$ | 245 | OM | HMs | 67 |
| $Mn_2CO_2$ | 110 | MX | HMs | 59 | $Fe_3GeTe_2$ | 216 | OM | Metal | 64-66 |
| $Mn_2C(OH)_2$ | 460 | MX | HMs | 59 | $Li_2N$ | 161 | OM | HMs | 60 |
| $Mn_2CCl_2$ | 380 | MX | HMs | 59 | $Fe_2Si$ | 780 | OM | HMs | 61 |
| $Mn_2CH_2$ | 120 | MX | HMs | 59 | $Cu_2MoS_4$ | 350 | OM | SCs | 68 |
| $Fe_3C$ | 861 | MX | Metal | 77 | $K_2CuF_4$ | 8 | OM | Insulator | 63 |
| $Cr_2Ge_2Te_6$ | 30 | TMT | SCs | 3 | $CrN$ | 675 | OM | HMs | 70 |
| $CrI_3$ | 45 | TH | SCs | 4 | $Co_2S_2$ | 404 | OM | Metal | 69 |
| $VSe_2$ | 300 | TMD | Metal | 5 | $Mn_3O_4$ | 45 | OM | -- | 71 |

We now turn to the status of experimental observations of ferromagnetism in 2D materials. Various magnetic materials have recently been explored to see whether their magnetism can be retained in their reduced dimensions, especially in their monolayers[42, 51, 63]. A few early reports indicated FM in few-layer $Cr_2Ge_2Te_6$ and anti-FM in monolayer $FePS_3$[78, 79]. However, the first clear observations of long-range intrinsic ferromagnetism in 2D systems were recently reported by two different groups revealing the ferromagnetic order in $CrI_3$ monolayer and $Cr_2Ge_2Te_6$ bilayer flakes at low temperatures[3, 4].

Huang and colleagues demonstrated the intrinsic FM in a mono layer of $CrI_3$ with a suppressed $T_c$ of 45 K[4], which they explain based on Ising exchange interaction. Interestingly they also observe a strong dependence of the ferromagnetism in $CrI_3$ on the number of layers as can be seen in figure 2. The observed remnant magnetization in a single layer is absent in a bilayer, indicating antiferromagnetism i.e. the two layers having oppositely oriented spins. Interestingly, the net magnetization is again recovered in a trilayer. This indicates the strong layer-dependent interplay between different mechanisms that stabilize magnetic ordering in the atomically thin limit.

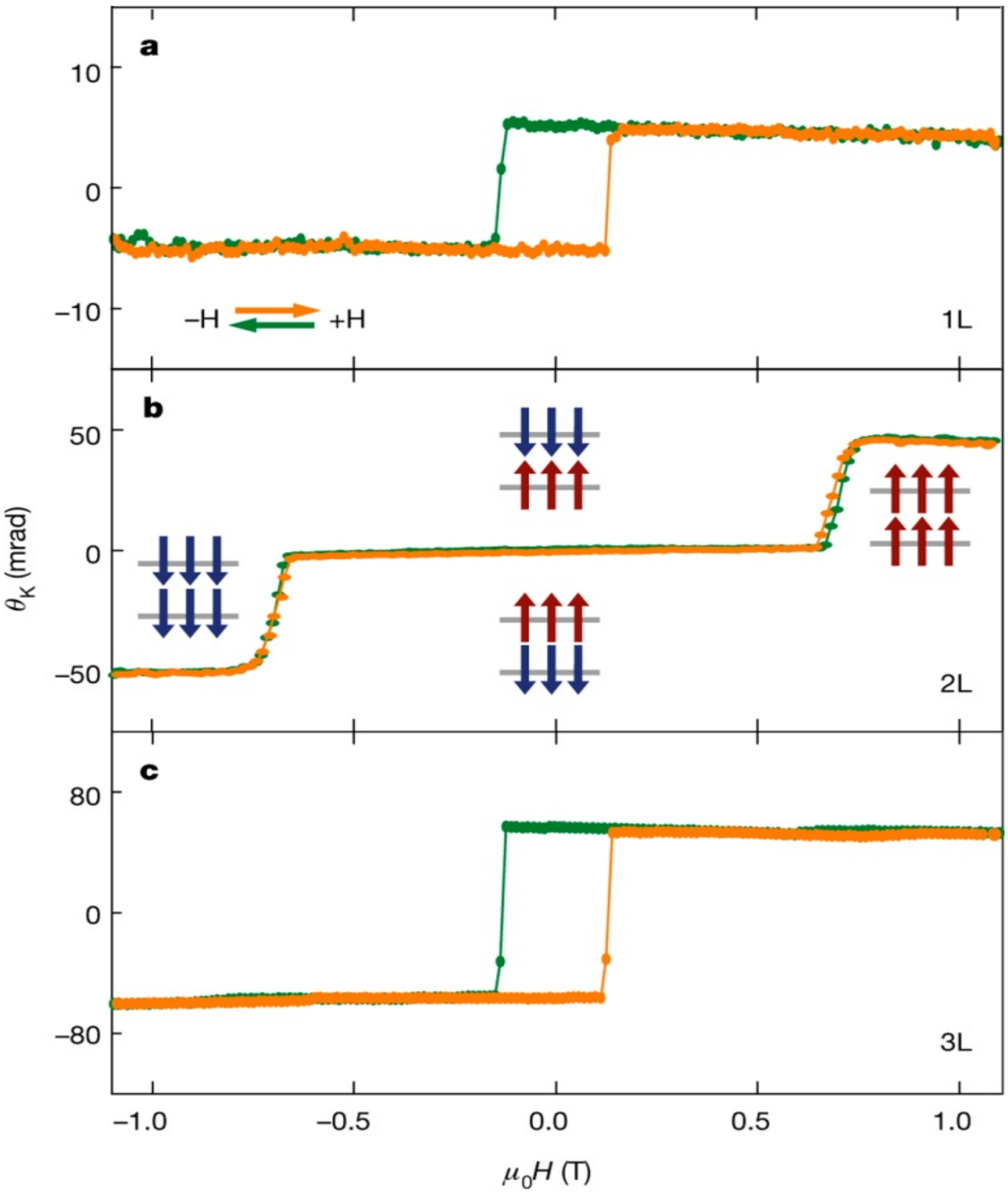

**Figure 2:** Layer-dependent magnetic ordering in atomically-thin $CrI_3$ [4]. **(a)** MOKE signal on a monolayer (1L) $CrI_3$ flake, showing hysteresis in the Kerr rotation as a function of applied magnetic field, indicative of ferromagnetic behavior. **(b)** MOKE signal from a bilayer $CrI_3$ showing vanishing Kerr rotation for applied fields $\pm$ 0.65 T, suggesting antiferromagnetic behavior. Insets depict bilayer (2L) magnetic ground states for different applied fields. **(c)** MOKE signal on a trilayer (3L) flake, showing a return to ferromagnetic behavior. Reprinted with permission from B. Huang et al., Nature 546 (7657), 270-273 (2017). Copyright 2018 Springer Nature Limited.

Gong and colleagues studied ultrathin layers of $Cr_2Ge_2Te_6$, demonstrating long range FMO also with a low $T_c$ ~ 30K in a bilayer[3]. The magnetic behavior in $Cr_2Ge_2Te_6$ can be explained by the Heisenberg model with anisotropic coupling, in which spins themselves are not direction dependent. The suppression of the material's Curie temperature with decreasing number of layers is closely linked to the thermal excitation of spin waves. Another relevant study about the anomalous paramagnetic to ferromagnetic transition in $CrGeTe_3$ is also carried out[80].

The intrinsic 2D ferromagnetism in $CrI_3$ and $Cr_2Ge_2Te_6$ is observed only far below room temperature (i.e. <293 K) and a challenge remained to realize room temperature intrinsic 2D ferromagnetism until the very recent remarkable discovery of the strong ferromagnetic ordering in monolayer $VSe_2$[5]. Notably, the FM in $VSe_2$ monolayer with a large magnetic moment persists to above room temperature, which is a major breakthrough in this field and that could lead to potential spintronic devices. Magnetic properties of $VSe_2$ are illustrated in figure 3.

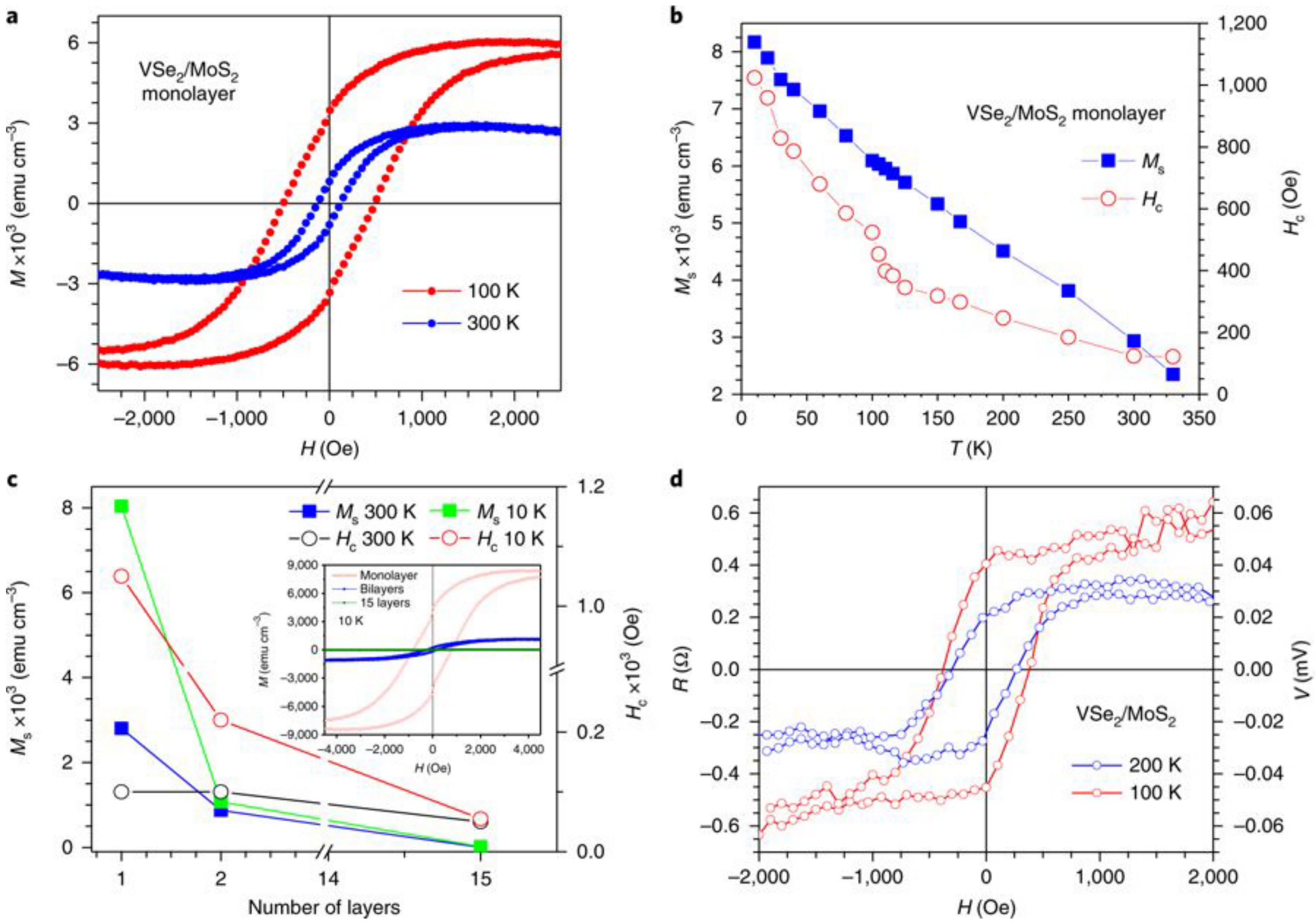

**Figure 3:** Magnetic properties of $VSe_2$ films on $MoS_2$ substrates[5]. **(a)** M–H loops took at 100 K and 300 K for monolayer $VSe_2$. **(b)** The strong temperature dependences of $H_c$ and $M_s$. **(c)** Variations of $M_s$ and $H_c$ with the number of layers of $VSe_2$ film. The inset shows the M–H loops for the mono-, bi- and multilayer samples. **(d)** Anomalous Hall-effect measurements. Magnetic field dependences of resistance (R) and voltage (V) taken at 100 K and 200 K show clear hysteresis with larger loops at a lower temperature (100 K versus 200 K), consistent with the temperature dependence of M–H loops observed by VSM. Reprinted with permission from M. Bonilla et al., Nature Nanotechnology 13, 289–293 (2018). Copyright 2018 Springer Nature Limited.

Contrary to 2D ferromagnetic materials $Cr_2Ge_2Te_6$ and $CrI_3$, long rang ferromagnetic ordering persists in monolayer $VSe_2$ with numerous unique features: (1) While ultrathin layers of $Cr_2Ge_2Te_6$ and $CrI_3$ are ferromagnetic insulators, monolayer of $VSe_2$ exists in 1T metallic phase with a trademark of CDW transitions. (2) While ultrathin layers of $Cr_2Ge_2Te_6$ and $CrI_3$ are cleaved from bulk ferromagnetic layered materials, monolayer $VSe_2$ exhibits ferromagnetic ordering even though the 3D bulk is paramagnetic which commonly condenses in the 1T structure. (3) Finally, the subject which is under heated discussion currently, ferromagnetic ordering in $VSe_2$ monolayer on graphite (highly oriented pyrolytic graphite (HOPG) or $MoS_2$ is possibly arising from the substrate effects.

The experimental demonstrations by Bonilla et al., showed that (i) the monolayer of $VSe_2$ cleaved from a $VSe_2$ single crystal has no ferromagnetic ordering. (ii) Measurements on HOPG and $MoS_2$ substrate without a $VSe_2$ film show a diamagnetic and weak defect-induced ferromagnetic background respectively. (iii) Both diamagnetic and weak ferromagnetic substrate's background cannot explain the large in-plane magnetization value, approximately 15 $\mu_B$ per unit cell, of $VSe_2$ monolayer. (iv) $M_s$ and $H_c$ for $VSe_2$ monolayer are strongly temperature dependent as shown in figure 3(b). These results suggest that ferromagnetic ordering is intrinsic property of $VSe_2$ monolayer and give clear indication that the ferromagnetic ordering in $VSe_2$ monolayer /HOPG or $VSe_2$ monolayer /$MoS_2$ sample is possibly stabilized from the substrate effects.

# IV. ROLE OF FERROMAGNETISM IN TOPOLOGICAL PHASES

Although Landau theory of symmetry or symmetry breaking cannot describe all the phases of matter – different phases of fractional quantum Hall effect possess same symmetries and their existence do not require breaking of any symmetries[81-88] – the symmetries of the underlying Hamiltonian and the breaking of certain symmetries still play an important role in describing and predicting exotic topological phases of quantum matter. For example, topological insulators (TI)[89-106] and Dirac semimetals[107-120] are protected through symmetries while the quantum integer/anomalous Hall (QIH/QAH) states[25, 121-126] and Weyl semimetals[109, 127-141] exist only if certain symmetries are broken.

The inter-winding among symmetries of the Hamiltonian and the topology of the ground state of the system – to probe band topology[100] – has already been generalized for all 230 space groups[142]. We are focused here on the role of time-reversal symmetry (TRS) and the long-range ferromagnetic order associated with TRS breaking in two-dimensions. Several topological phase transitions, *i.e.*, from Wallace (Semenoff) topologically trivial semi-metallic (insulating) phase[143, 144] to Haldane topological Chern insulating phase [25], from Dirac to Weyl semi-metal phase[128-130], and from two-dimensional topological insulating phase[89-106, 145] to QAH phase [122-126], require breaking of TRS. Hence ferromagnetic materials, or heterostructures containing them, are natural candidates for realizing these phases in the absence of an applied magnetic field.

Such a phase transition led by magnetic ordering associated with the breaking of TRS is not limited to topological insulators and semi-metals. For example, the interfacing among topological states with other conventional ordered states such as superconducting order has led to the prediction of many new exotic phases of quantum matter [100, 124, 146-164]. The presence of chiral edge states in both topological Chern insulator, as well as conventional spinless $p_x + ip_y$ superconductors, are associated with breaking of TRS. Fu and Kane showed that a two-dimensional state – the Fu-Kane superconductor[146] – that resembles spinless $p_x + ip_y$ superconductor which host both Majorana bound states associated with vortices[165-168] as well as chiral edge states [165, 169, 170] – can be obtained through proximity effect between an ordinary *s*-wave superconductor and the surface states of strong TI but without TRS breaking. However, to have chiral Majorana edge states in Fu and Kane model, TRS needs to be broken by introducing insulating FMO at the topological insulator-superconductor interface.

Meng and Balents have also shown a similarity between two-dimensional $p_x + ip_y$ superconductors and a Weyl superconductor – a superconducting phase emerged from the normal insulator, topological insulator, and superconductor heterostructure[171, 172]. The consequence of TRS and its breaking are similar to what one finds in the phase transition from a trivial superconducting phase to a Fu-Kane superconducting phase: A time reversal symmetric (non-centrosymmetric-inversion broken) Weyl superconductor phase is trivial while a centrosymmetric (time reversal breaking) Weyl superconductor phase is topologically non-trivial superconductor with chiral Majorana surface modes.

In short, many phase transitions from topologically trivial semi metallic, insulating, or superconducting phases to topologically non-trivial semi metallic, insulating, or superconducting phases require breaking of TRS. These daughter topological non-trivial phases allow possible future dissipationless spintronic and electronic devices by achieving long-range ferromagnetic order in parent topologically trivial/non-trivial phases.

Since it is not possible to cover all the phase transitions based on TRS breaking in this focused review, we highlight here the possible realization of quantum anomalous Hall effect through breaking of TRS in TIs as the future of dissipationless transport is inherited in the experimental realization of quantum anomalous Hall effect. To grasp the idea, we review the basic models for topological insulating phases, the origin of ferromagnetic order in TIs, and finally the realization of QAH phase with the emergence of long range ferromagnetic order in TIs. Surface Dirac cones and the bulk band structure of three-dimensional topological insulator doped with TR invariant and TR breaking impurities are shown in figure (4).

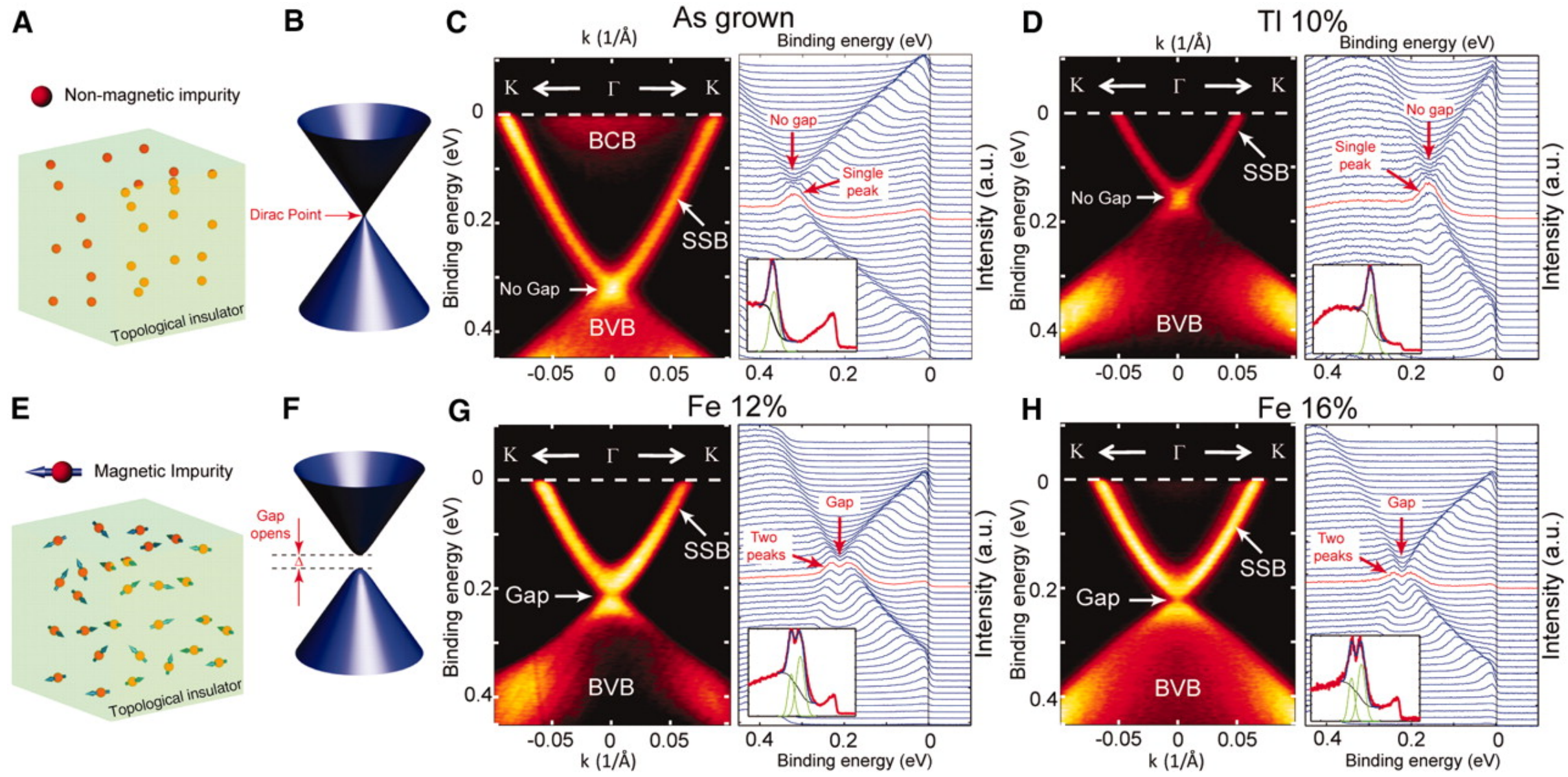

**Figure 4:** (**A** and **B**) A topological insulator doped with TR-invariant impurities hosts a Dirac point similar to the one in the undoped case. (**C and D**) Band structure along the K-Γ-K direction of undoped $Bi_2Se_3$ and Tl-doped $(Bi_{0.9}Tl_{0.1})_2Se_3$ respectively. The Dirac point remains gapless in both cases. (**E** and **F**) A topological insulator doped with TR-breaking impurities hosts a broken Dirac point. (**G** and **H**) Band structure of two Fe-doped TIs $(Bi_{0.88}Fe_{0.12})_2Se_{3.7}$ and $(Bi_{0.84}Fe_{0.16})_2Se_{3.7}$ respectively. A gap formation occurs at the Dirac point in both cases[145]. Reprinted with permission from Y. L. Chen et al., Science 329 (5992), 659-662 (2010). Copyrights 2018 American Association for the Advancement of Science.

## V. TOPOLOGICAL INSULATING PHASE

To show how TRS and its breaking are associated with the TI phase and the QAH phase respectively, we discuss here a four-band model which serves as a guiding principle for generating QAH phase in magnetically doped TIs[122-126]. In a suitably chosen basis, the time reversal invariant four-band mode model written as

$$H = \begin{pmatrix} h(\mathbf{k}) & 0 \\ 0 & h^*(-\mathbf{k}) \end{pmatrix} \qquad (2)$$

where the two-band component blocks $h(\mathbf{k})$ and $h^*(-\mathbf{k})$ are TR partners. Such a model describes both the TI phase found in semiconducting quantum wells (QW) as well as the TI phase found in two-dimensional thin films of tetradymite semiconducting family $Bi_2Se_3$, $Bi_2Te_3$, and $Sb_2Te_3$. For example, in the basis $\{|E_1 +\rangle, |H_1 +\rangle, |E_1 -\rangle, |H_1 -\rangle\}$ where $|E_1 \pm\rangle$ and $|H_1 \pm\rangle$ are the degenerate first electron sub-bands and heavy-hole sub-bands respectively, this four-band model represents TI phase in semiconducting QW where $h(\mathbf{k})$ takes the form

$$h(\mathbf{k}) \rightarrow h_{2D-TI}(\mathbf{k}) = C - D(k_x^2 + k_y^2) + A(k_x\sigma_x - k_y\sigma_y) + (M - B(k_x^2 + k_y^2))\sigma_z \qquad (3)$$

The material parameters *A*, *B*, *C*, *D*, and *M* depend on the QW geometry. This four-band model describes TI phase found in both type-I HgTe and type-II InAs/GaSb semiconducting quantum wells; hybridization gap generated through coupling *A* between electron and hole sub-bands in type-II QW is relatively smaller than that in type-I QW.

Similarly, this four-band model also represents the two-dimensional limit of 3D TI in the basis $|+\uparrow\rangle, |-\downarrow\rangle, |+\downarrow\rangle, |-\uparrow\rangle$ where $|\pm\uparrow\rangle = (|t\uparrow\rangle \pm |b\uparrow\rangle)/\sqrt{2}$, $|\pm\downarrow\rangle = (|t\downarrow\rangle \pm |b\downarrow\rangle)/\sqrt{2}$ and t (and b) represent the top (and bottom) surface states while ↑, ↓ represent the spin up and down states respectively. Here, $h(\mathbf{k})$ takes the form

$$h(\mathbf{k}) \rightarrow h_{3D-TI}^{surf}(\mathbf{k}) + m(\mathbf{k})\sigma_y \qquad (4)$$

where $m(\mathbf{k}) = M - B(k_x^2 + k_y^2)$ represents the tunneling effect between the top and bottom surface states at the surface of 3D TI with normal vector $\hat{z}$

$$h_{3D-TI}^{surf}(\mathbf{k}) = \varepsilon(\mathbf{k}) + v_f(k_y\sigma_x - k_x\sigma_y) \qquad (5)$$

Here $v_f$ is the Fermi velocity. Bulk energy bands of the type-I semiconducting quantum well are shown in figure (5) while the local density approximation for the band structure of three-dimensional bulk and the ARPES measurements on the two-dimensional surface (111) of $Bi_2Se_3$ is shown in figure (6).

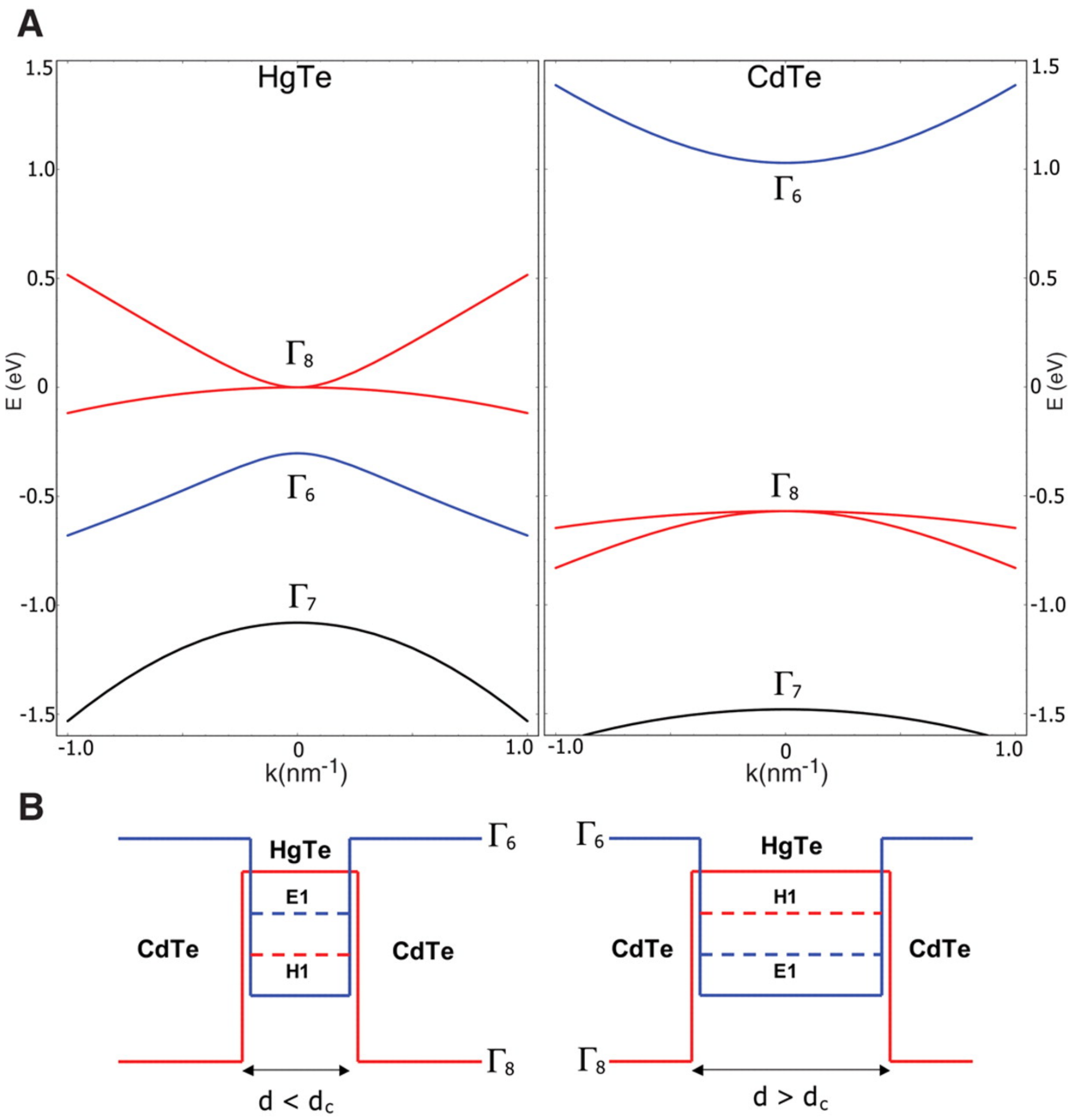

**Figure 5:** (**A**) Bulk energy bands of HgTe and CdTe near the Γ point. (**B**) The CdTe-HgTe-CdTe quantum well in the normal regime ($d < d_c$) and in the inverted regime ($d > d_c$)[91]. Reprinted with permission from B. A. Bernevig et al., Science 314 (5806), 1757-1761 (2006). Copyrights 2006 American Association for the Advancement of Science.

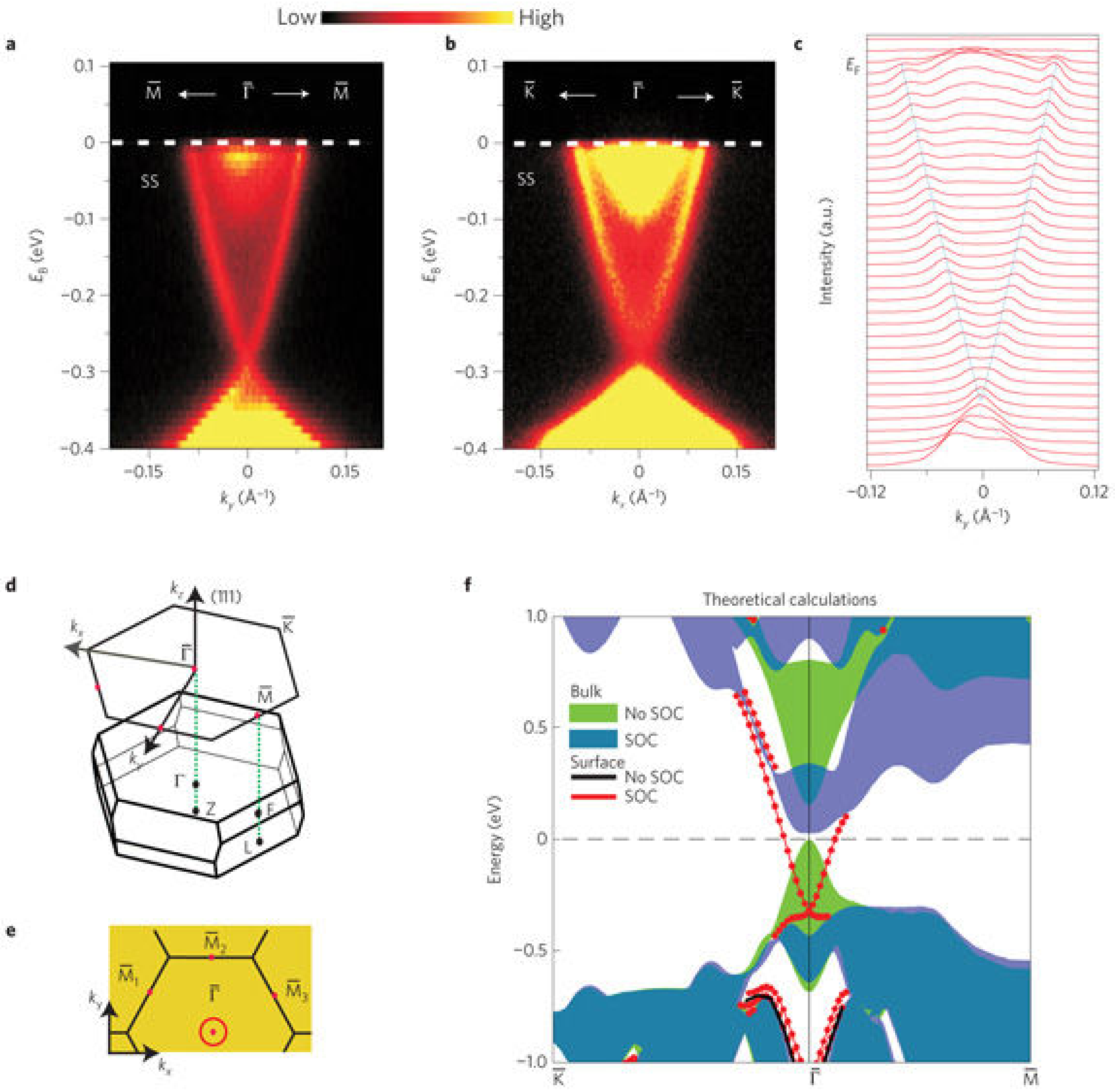

**Figure 6: (a, b)** Electronic band dispersion of $Bi_2Se_3$ through ARPES measurements on the surface (111) with an incident photon energy of 22 eV near the $\bar{\Gamma}$-point along the $\bar{\Gamma}-\bar{M}$ and $\bar{\Gamma}-\bar{K}$ momentum-space cuts respectively. **(c)** The corresponding momentum distribution curves. **(d)** A schematic diagram of the three-dimensional bulk and the two-dimensional surface (111) BZ of $Bi_2Se_3$. **(e)** When the chemical potential lies inside the bulk bandgap, a single ring centered at $\bar{\Gamma}$ represents the Fermi surface of the surface states on the two-dimensional surface (111) along the $\bar{K}-\bar{\Gamma}-\bar{M}$ momentum-space cut. The red dots at $\bar{\Gamma}$ and the three $\bar{M}$ points represent TRIMs on the surface (111) BZ. f) Local density approximation of the Bulk (represented by the shaded areas) and the surface (represented by the red dotted lines) band structure. In the presence of strong spin–orbit coupling, the existence of a π Berry phase on the surface is guaranteed since the surface band crosses the Fermi level only once between the TRIMs $\bar{\Gamma}$ and $\bar{M}$ [105]. Reprinted with permission from Y. Xia et al., Nature Physics 5, 398 (2009). Copyright 2009 Springer Nature Limited.

## VI. ORIGIN OF FERROMAGNETISM IN TOPOLOGICAL INSULATORS

At present, no intrinsically ferromagnetic topological 2D material is known. Ferromagnetism has successfully been introduced into topological materials through substitutional doping. In analogous dilute magnetic semiconductors, the magnetic order can be established in both metallic and insulating phases through the Ruderman-Kittel-Kasuya-Yosida (RKKY) mechanism and Van Vleck paramagnetism, respectively. In the conducting phase, exchange coupling between itinerant electron spin density and spin of magnetic impurity leads to the RKKY interaction among magnetic impurities $\mathbf{S}_1(\mathbf{r})$ and $\mathbf{S}_2(\mathbf{r}')$

$$H_{\text{int}}^{RKKY} = \sum_{i,j=x,y,z} J_{jk}(|\mathbf{r}-\mathbf{r}'|)\mathbf{S}_{1i}(\mathbf{r})\mathbf{S}_{2j}(\mathbf{r}') \qquad (6)$$

where $J_{jk}(|\mathbf{r}-\mathbf{r}'|)$ is the coupling constant between two magnetic impurities separated by $|\mathbf{r}-\mathbf{r}'|$. In the insulating phase, a non-vanishing second order perturbation term in the ground state energy leads to Van Vleck paramagnetic spin susceptibility for band insulator

$$\chi_z^{VV} = \frac{2\mu_0\mu_B^2}{V}\sum_{m\neq n}\frac{|\langle u_{n,k}|s_z|u_{m,k}\rangle|^2}{E_m - E_n} \qquad (7)$$

where $E_m > E_n$. Here $u_{m,k}$ and $u_{m,k}$ are the Bloch functions in valance and conduction band, $\mu_0$ is the vacuum permeability, $\mu_B$ is Bohr magneton, and $S_z$ is the electron spin operator.

In magnetically doped 2D TIs found in type-II semiconducting quantum wells InAs/GaSb[95], as well as thin films of 3D TIs, found in magnetic tetradymite semiconducting family $Bi_2Se_3$[173], both RKKY interaction as well as Van Vleck paramagnetic electron susceptibility can contribute in ferromagnetic order. However, due to large hybridization gap in type-I semiconducting quantum wells HgTe, the ferromagnetic order is hard to establish. Although magnetically doped 2D TIs in type-II semiconducting quantum wells has not been realized experimentally, the presence of both RKKY ferromagnetic order and finite Van Vleck paramagnetic contribution has led to the experimental realization of long range ferromagnetic order in 2D thin films of dilute tetradymite semiconductors family $Bi_2Se_3$, $Bi_2Te_3$ , and $Sb_2Te_3$ when doped with transition metal elements such as Fe[174], Cr[175], and V[176]. The Van Vleck type spin susceptibility of $Bi_2Se_3$ bulk as a function of the SOC strength is shown in figure (7)

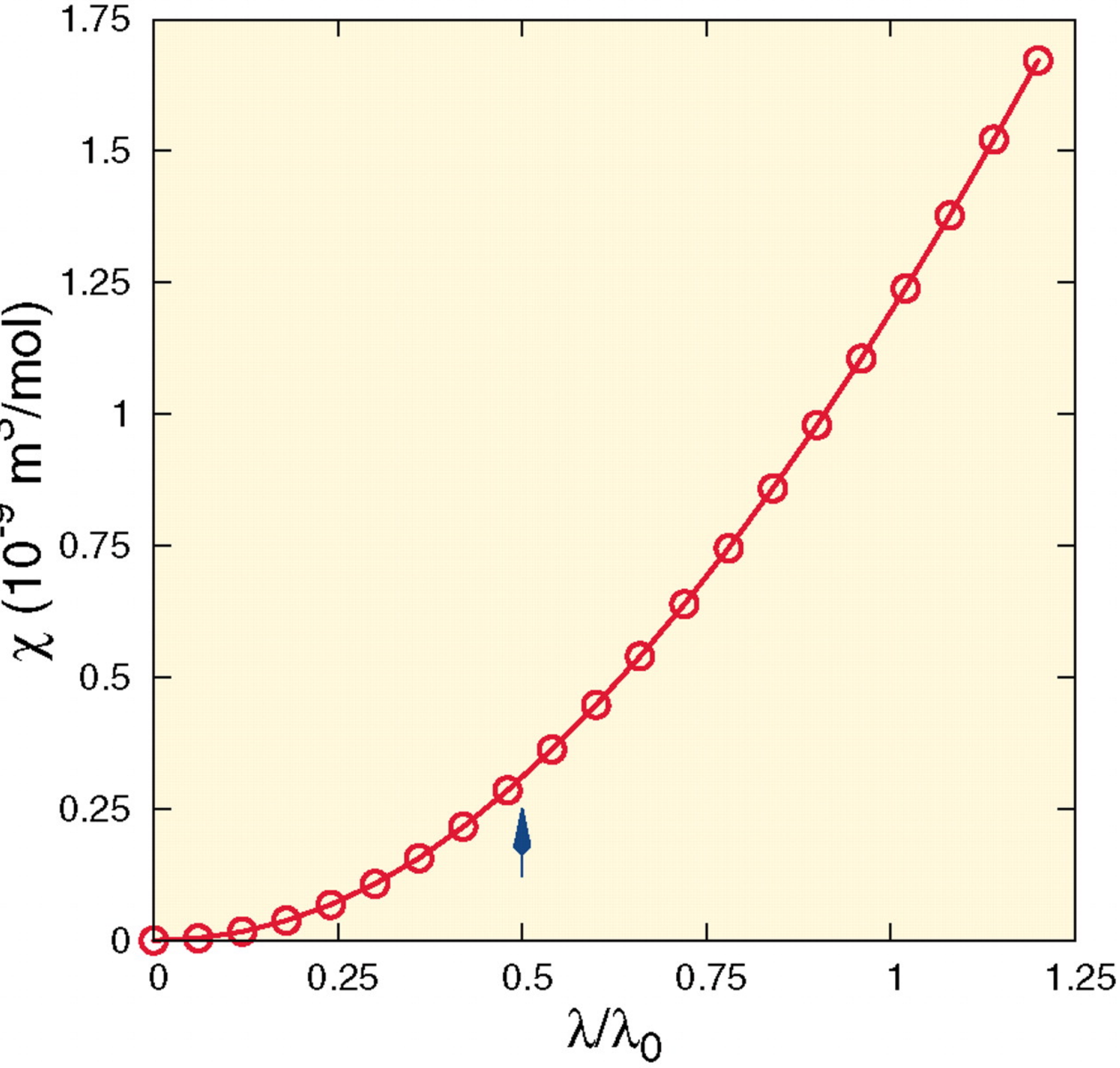

**Figure 7:** The Van Vleck paramagnetic electron spin susceptibility of $Bi_2Se_3$ bulk as a function of the SOC strength. Here $\lambda_0$ represents the actual SOC strength[125]. Reprinted with permission from R. Yu et al., Science 329 (5987), 61-64 (2010). Copyrights 2010 American Association for the Advancement of Science.

## VII. IMPORTANCE OF FERROMAGNETISM IN REALIZATION OF "QAH" PHASES

The QAH insulating phase can be realized in systems with spontaneous magnetization and/or strong spin-orbit-interaction. The presence of both collectively gives rise to a topologically non-trivial band structure characterized by non-zero Chern number. Since the observation of the TI phase in 2D semiconducting QW and 3D tetradymite semiconductors is the consequence of strong SOC, the key ingredient to observe QAH phase is to have long range ferromagnetic order in TIs.

The role of a ferromagnetic order associated with TR symmetry breaking can be formulated in terms of TR breaking four-band mode model as

$$\widetilde{H} = \begin{pmatrix} h(\mathbf{k}) + h_{FMO} & 0 \\ 0 & h^*(-\mathbf{k}) - h_{FMO} \end{pmatrix} \quad (8)$$

where $h_{FMO}$ represents the conventional ferromagnetic order. To generate QAH insulating phase with chiral edge states, $h_{FMO}$ must fulfill the following two constraints such that the TR partner 2-band component blocks have (i) different mass terms to assure that TR symmetry in broken and (ii) different band structure; inverted band structure for one block while normal/non-inverted band structure in the other block.

For experimental realization, such ferromagnetic order can be achieved by introducing magnetic exchange coupling either through magnetic proximity effect – by combining topological trivial/non-trivial parent material with a topological trivial magnetic material such that the spin moments of band/iterant electrons in parent material are locally aligned out of plane – or through inducing magnetic impurities in the parent material. With a suitable choice of Zeeman or magnetic exchange coupling term in TR breaking the four-band model $\widetilde{H}$, TI phase found in both semiconducting QWs as well as two-dimensional thin films of tetradymite semiconducting family could lead to QAH phase.

Originally, Liu et al. proposed the QAH insulating phase in Mn doped 2D TI found in type-I semiconducting quantum wells such as $Hg_{1-x}Mn_xTe$ QW where FMO takes the form

$$h_{FMO} = G_+ I + G_- \sigma_z \quad (9)$$

Here $G_\pm = (G_1 \pm G_2)/2$ and $G_1$ and $G_2$ induce spin splitting in electron sub-bands $|E_1, \pm\rangle$ and heavy hole sub-bands $|H_1, \pm\rangle$ respectively. To fulfill the above-mentioned constraints on $h_{FMO}$ and achieve the QAH phase in $Hg_{1-x}Mn_xTe$ QW, $G_1$ and $G_2$ must be such that $G_1 \neq G_2$ and $G_1 G_2 < 0$.

However, QAH phase in $Hg_{1-x}Mn_xTe$ QW system could not be observed because Mn moments do not order spontaneously and hence cannot generate required spin splitting energy.

As a result, QAH phase is hard to realize experimentally without applying an external magnetic field in a 2D magnetic TI like $Hg_{1-x}Mn_xTe$ QW. To overcome this problem, Yu et al[125] predicted that QAH phase could be realizable in two-dimensional thin films of magnetically doped 3D TI found in tetradymite semiconductors $Bi_2Te_3$, $Bi_2Se_3$, and $Sb_2Te_3$. In this case, the ferromagnetic order takes the form

$$h_{FMO} = gm\sigma_z \qquad (10)$$

where the effective $g$ factor and magnetization along the z-axis $m$ take the same value for top and bottom surfaces. The constraint $|m| > |M|$ fulfills both the conditions on $h_{FMO}$ and hence the QAH phase is guaranteed. This prediction led to the experimental realization of QAH phase in thin films of Cr-doped $(Bi,Sb)_2Te_3$[177]. Magnetic field dependence of $\rho_{yx}$ and $\rho_{xx}$ for $Cr_{0.15}(Bi_{0.1}Sb_{0.9})_{1.85}Te_3$ film at different temperatures and the different gate voltage is shown in figure (8,9).

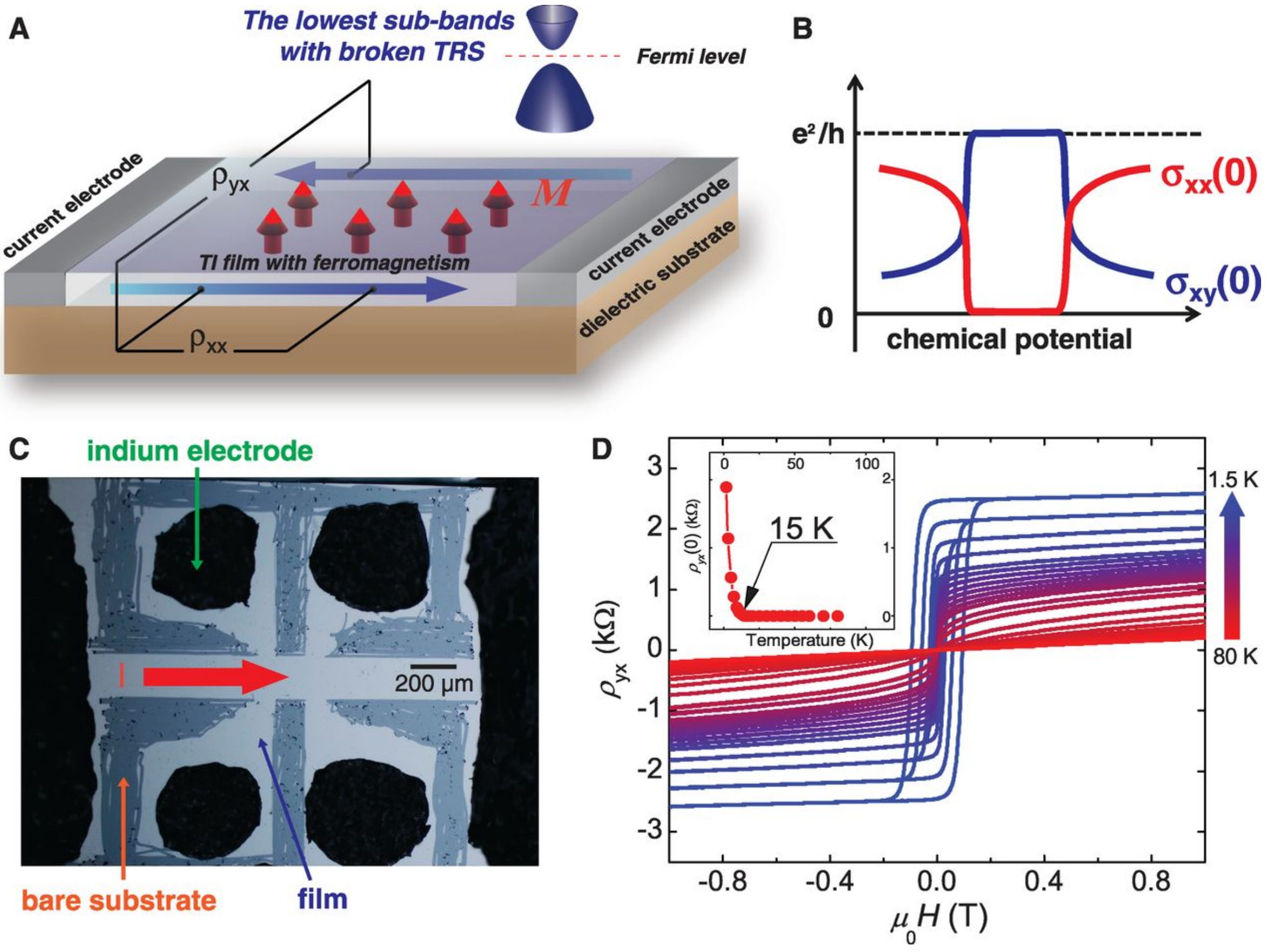

**Figure 8:** (**A**) A schematic representation for the realization of the QAH effect in a two-dimensionally thin film of magnetically doped TI. The chemical potential of the film can be controlled by applying gate voltage to a dielectric substrate. (**B**) A schematic drawing showing the expected dependence of zero field conductivities $\sigma_{xx}(0)$

and $\sigma_{xy}(0)$ on chemical potential. (**C**) An optical image of a Hall bar device where the light gray areas are the $Cr_{0.15}(Bi_{0.1}Sb_{0.9})_{1.85}Te_3$ film, dark gray areas are bare substrate from where the film is removed, the red arrow indicates the direction of current flow during the measurements, and the black areas are the attached indium electrodes. (**D**) Magnetic field dependence of $\rho_{yx}$ for $Cr_{0.15}(Bi_{0.1}Sb_{0.9})_{1.85}Te_3$ film measured at different temperatures and the temperature dependence of zero field $\rho_{yx}(0)$ shown in the ansatz which indicates a Curie temperature of 15 K[177]. Reprinted with permission from C.-Z. Chang et al., Science 340 (6129), 167-170 (2013). Copyrights 2013 American Association for the Advancement of Science.

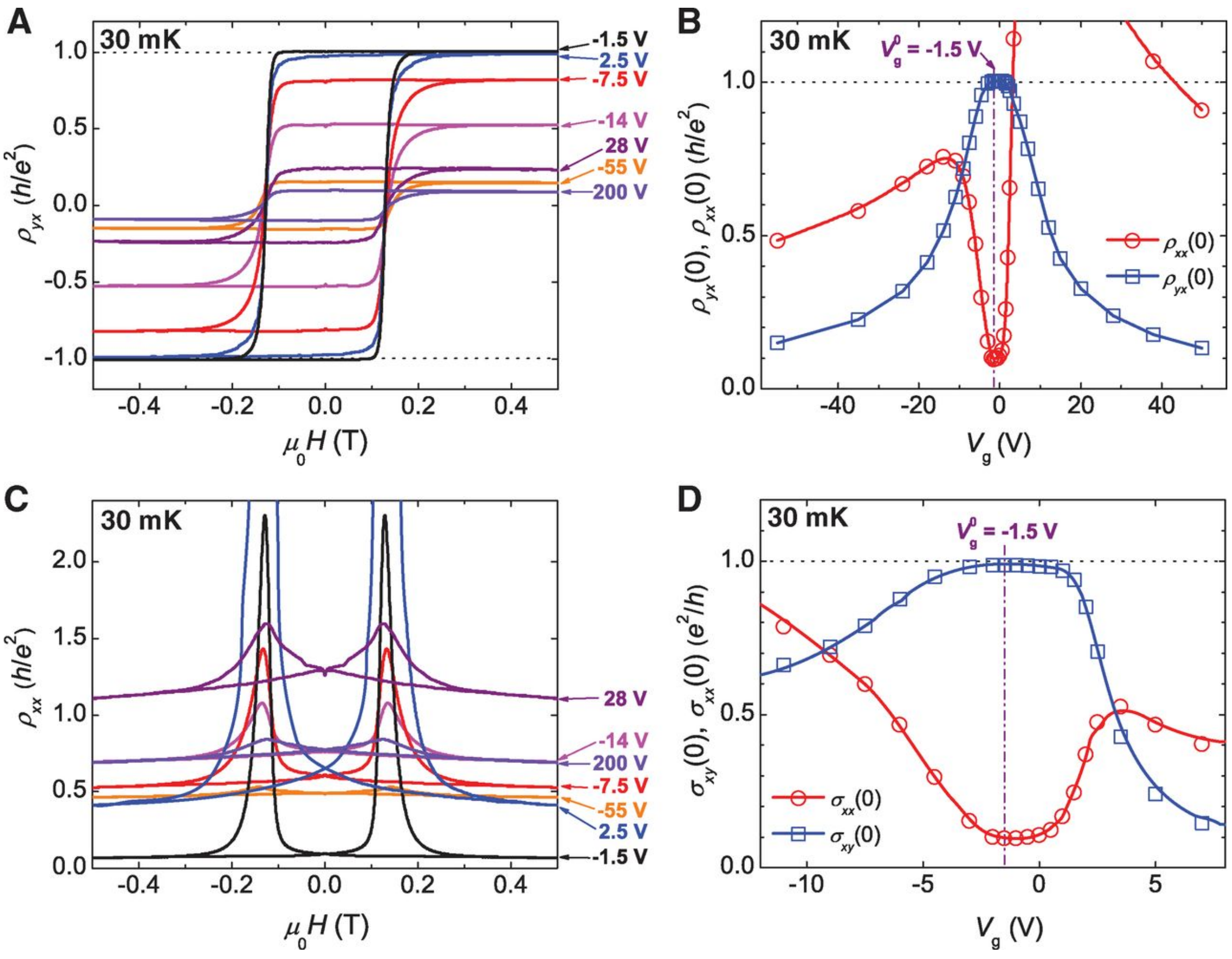

**Figure 9:** (**A** and **C**) Magnetic field dependence of $\rho_{yx}$ and $\rho_{xx}$ at different gate voltage applied to dielectric substrate $V_g$. **(B and D)** $V_g$ dependence of zero field resistivities $\rho_{yx}(0)$, $\rho_{xx}(0)$ and zero field conductivities $\sigma_{xy}(0)$ and $\sigma_{xx}(0)$[177]. Reprinted with permission from C.-Z. Chang et al., Science 340 (6129), 167-170 (2013). Copyrights 2013 American Association for the Advancement of Science.

Ideally, the QAHE would manifest in exactly zero $\rho_{xx}$ at zero magnetic field, corresponding to the dissipationless transport of charge along the QAH edges. However, the presence of spin scattering centers due to non-uniform magnetization and parallel dissipative channels at the edge of Cr-doped $(Bi,Sb)_2Te_3$[177] likely induces several KΩ longitudinal resistance in the absence of an external magnetic field. Hence the ideal QAH phase with dissipationless edge conduction in Cr-doped $(Bi,Sb)_2Te_3$[177] still requires an external magnetic field of several teslas. Later, both of these mechanisms responsible for dissipation are addressed in V-doped[178] and Cr-doped[179] tetradymite family $(Bi,Sb)_2Te_3$.

Moreover, difficulty in achieving both a high $T_c$ and a large band gap opening simultaneously in magnetically doped TIs limits the applications of QAHE at room temperature. Room-temperature ferromagnetism is the key requirement for the realization of QAHE and its utilization in related spintronics and low-energy electronics technologies. In the QAH phase, quantized transverse resistance and extremely low longitudinal resistance allow possible future dissipationless spintronic and electronic devices. However, the highest temperature at which QAH phase is observed so far is less than 100 mK. For practical applications, the prime goal is to achieve dissipationless chiral edge states at room temperature. To achieve this, the key factor is intrinsic ferromagnetic order sustainable against thermal fluctuations at high temperature in the two-dimensional limit. In short, the discovery of intrinsic ferromagnetic order with high $T_C$ in atomically thin materials with topological non-trivial band structure would offer a much better platform to realize the QAHE without random magnetic doping.

## VIII. FUTURE DIRECTIONS

The discovery of intrinsic 2D ferromagnetism could revolutionize modern day technologies in many ways from both comprehending the underlying physics of electronic and spin processes, and applications ranging from sensing to information, and communication technologies[180]. If 2D magnetic semiconductors are the building blocks for modern sensing and communication technologies, ferromagnetic metals with reduced dimensionality on the other hand could help us to develop fast and non volatile storage/memory devices purely based on the fundamentals of physics where not only the charge transport is controlled by the magnetization orientation but the magnetization orientation can be electrically modulated[181-183].

Intrinsic ferromagnetism in 2D materials also plays an essential role in realizing new ultra-low energy electronics in several ways. Firstly, strong spin-orbit coupling in thin film materials along with surface and interface effects are important in many spin-based phenomena and low-energy spintronic applications. Secondly, topologically protected spin-polarized surface states of topological insulators provide efficient conversion from spin to charge transport and vice versa. This spin-charge conversion is greatly enhanced with the reduction of dimension. Finally, the intrinsic ferromagnetic order in two-dimensionally thin films of topological insulators opens the possibility of dissipationless charge transport along the edges of the quantum anomalous Hall phase.

The integration of semiconductor spintronic devices with ferromagnetism could also be useful in quantum information applications that are based on manipulation of spin states in semiconductors[184-186]. Recent discoveries of intrinsic 2D ferromagnetism could potentially lead us to discover new spin-gapless semiconductors (SGS)[187] which link semiconductors and half-metals. The idea of the SGS was first coined by Wang et al. in 2008[188] based on the novel designs of band structures with either linear or parabolic energy dispersion. The SGSs with linear energy dispersion are promising materials for realizing massless and dissipationless states.

In summary, the discovery of intrinsic ferromagnetism in two-dimensionally thin materials with signatures of dissipationless spintronic and electronic transport urges the discovery of new spin-and-charge based phenomena where robustness against disorder and thermal fluctuations promises spintronic applications at room temperature. Intrinsic ferromagnetism in combination with recent thin film growth techniques provides novel directions to synthesize and fabricate possible candidates to observe high temperature QAHE phenomena for electronics and related spintronics applications.

**Additional Note:** The experimental demonstration of intrinsic ferromagnetism in $VSe_2$ monolayer /HOPG or $VSe_2$ monolayer /$MoS_2$ sample in 1T metallic phase and CDW transitions triggered a heated debate on substrate-induced ferromagnetic ordering and CDW transitions. Numerous articles have appeared on long range ferromagnetism and CDW phase maintainable on substrates since drafting this manuscript and its consideration. For example, transition metal compound monolayers such as 1T-$MnSe_2$[189, 190], $Fe_3GeTe_2$ [191], and transition metal dihydride $MH_2$ (M=Sc, Co)[192] host magnetic states and the van der Waals substrate stabilizes the former and leads toward 2D itinerant ferromagnetism. More interestingly, the hetro-interface plays key role in deriving a metal-to-insulator transition in the TMDC monolayers by inducing a band gap up several meV and leads to CDW phase in transition metal compounds[193, 194]. This discussion has also been extended to electric field controlled (Gate-tuning effects on) magnetism in a few layered van der Waals ferromagnetic $Cr_2Ge_2Te_6$ [195], and antiferromagnetic $CrI_3$ semiconductors[196, 197].

**Acknowledgments**

B.S. and M.N. contributed equally to this work. B.S. acknowledges the funding support from China Postdoctoral Science Foundation Grant (No. 2017M622758). M.N. acknowledges support from the ARC Professional Future Fellowship (FT130100778). Z.D. acknowledges the National Natural Science Foundation of China Grant (No. 51601131)，International Postdoctoral Exchange Fellowship Program of China, and Hubei Provincial Natural Science Foundation (2016CFB166).